# Synthesis and characterization of chloride doped polyaniline by bulk oxidative chemical polymerization Doping effects on electrical conductivity

Yomen Atassi, Mohammad Tally and Mazen Ismail

Higher Institute for Applied Sciences and Technology, HIAST, P.O. Box 31983, Damascus, Syria.

Abstract: Conductive polymers or "organic metals" are highly engineered nanostructured materials made from organic building blocks. They are candidates as molecular wires for nanotechnology applications in molecular electronics. The conduction in these polymers is due to the presence of delocalized molecular orbitals. In this work, we present the synthesis of chloride doped polyaniline by bulk oxidative chemical polymerization using a solid aniline salt as a monomer instead of liquid aniline to diminish toxic hazards. The FTIR and UV-visible spectra confirmed the expected structure of the polymer. The electrical conductivity measured using a four-probe method was 1.7 S/cm. The dependence of impedance modulus on frequency was measured using an HP impedance analyzer in the range 10 kHz-13 MHz. The influence of doping and the preparation temperature on the electrical conductivity were also investigated.

Corresponding author e-mail address, Yomen_Atassi@hiast.edu.sy

1. **Introduction**

Conducting polymers have the electronic properties of semiconductors and, at the same time, the processing advantages and mechanical parameters of polymers. Conducting polymers are characterized by a conjugated structure of alternating single and double bonds. The feature shared by all them originates from the common nature of their $\pi$ – electron system, an enhanced conductivity in oxidized or in reduced state and reversible redox activation in a suitable environment [1].

The fundamental process of doping is a charge-transfer reaction between an organic polymer and a dopant. When charges are removed from (or added to) a polymer upon chemical doping, geometric parameters, such as bond length and angles, change. The charge is localized over the region of several repeating units. Since the localized charges can move along the polymer chain, they are regarded as charge carriers in conducting polymer. These quasi-particles are classified into polarons, bipolarons and solitons according to their charge and spin [2].

Conducting polymers have been found suitable for microelectronic device fabrication due to their excellent electric characteristics and ease of processability. Among these polymers, polyaniline (PANI) has emerged as a promising candidate with great potential for practical uses such as in light emitting diodes, transparent electrodes, electromagnetic radiation shielding, corrosion protection of metals, gas and humidity sensing, battery applications and many more.

Usually polyaniline is made conducting via protonic acid doping e.g. HCl, or via oxidation either chemically or electrochemically [3].
Polyaniline exists in a variety of protonation and oxidation forms. The most important form of polyaninline, green protonated emeraldine (Fig.1) is produced by the oxidative polymerization of aniline in aqueous acids. It's electrically conducting (conductivity $\sigma = 10^{-2} - 10^{0} \text{S/cm}$), due to the presence of cation radicals in its structure. The positive charge on aniline units is balanced by negatively charged counterions, typically chloride or sulfate anions.

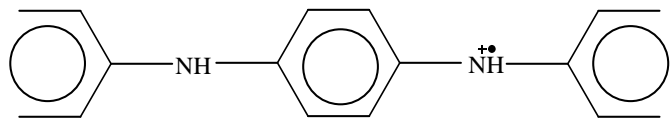

Fig.1: Green protonated emeraldine.

As the efficient polymerization of aniline is achieved only in an acidic medium, where aniline exists as an anilinium cation, we have selected hydrochloric acid in equimolar proportion to aniline, i.e., aniline hydrochloride as a monomer [4]. Ammonium peroxydisulfate was used as an oxidant. To minimize the presence of residual aniline and to obtain the best yield of PANI, the stoichiometric peroxydisulfate/aniline ratio was 1.25. The oxidation of aniline hydrochloride with ammonium peroxydisulfate to yield polyaniline (emeraldine) hydrochloride is presented in Fig.2.

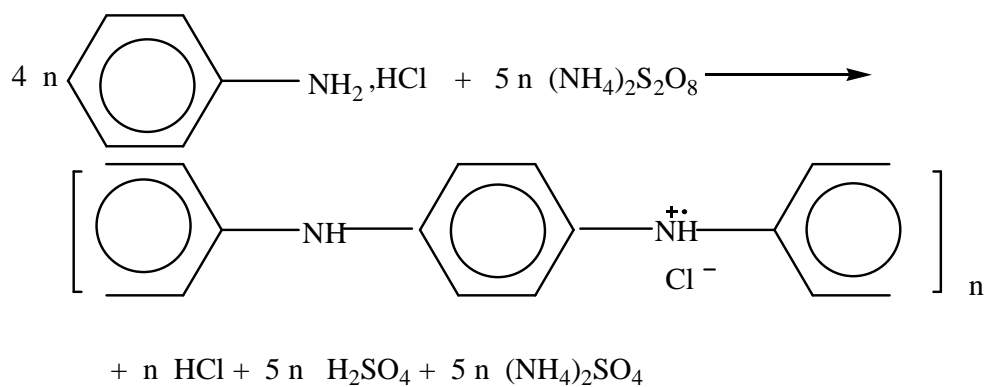

Fig.2: Oxidation of aniline hydrochloride with ammonium peroxydisulfate to yield polyaniline (emeraldine) hydrochloride.

## 2. Experimental Section
### 2.1. Polymer preparation
- **polyaniline hydrochloride (PANI salt)**

The synthesis was based on mixing aqueous solutions of aniline hydrochloride 0.2M and ammonium peroxydisulfate 0.25M at room temperature, followed by the separation of PANI hydrochloride precipitate by filtration and drying. More precisely, 2.59 g of anilinium hydrochloride was dissolved in distilled water in a volumetric flask to 50 mL of solution. 5.71 g of ammonium peroxydisulfate was dissolved in water also to 50 mL of solution. Both solutions were kept for 1 h at room temperature ($\sim 18 - 24°C$), then mixed in a beaker, briefly stirred, and left at rest to polymerize. Next day, the resulting dark green PANI precipitate was collected on a filter, washed with three 100 mL portions of 0.2 M HCl, and similarly with acetone. Polyaniline (emeraldine) hydrochloride powder was dried in air and then in vacuo at $60°C$. The polymerization is completed within 10 min at room temperature and within 1 h at 0-2$^0$C. The oxidation of aniline is exothermic so the temperature of the reaction mixture can be used to monitor the progress of reaction. Fig.3 illustrates the temperature profile in the polymerization of aniline.

After a brief induction period, polymerization begins and the temperature of the mixture increases; it passes through a maximum after the reaction is finished, and the medium cools down. So, using large volumes of high concentration aniline solution can result in the overheating of the system, and may lead to an explosion. A lot of precautions should be taken in such situations [4].

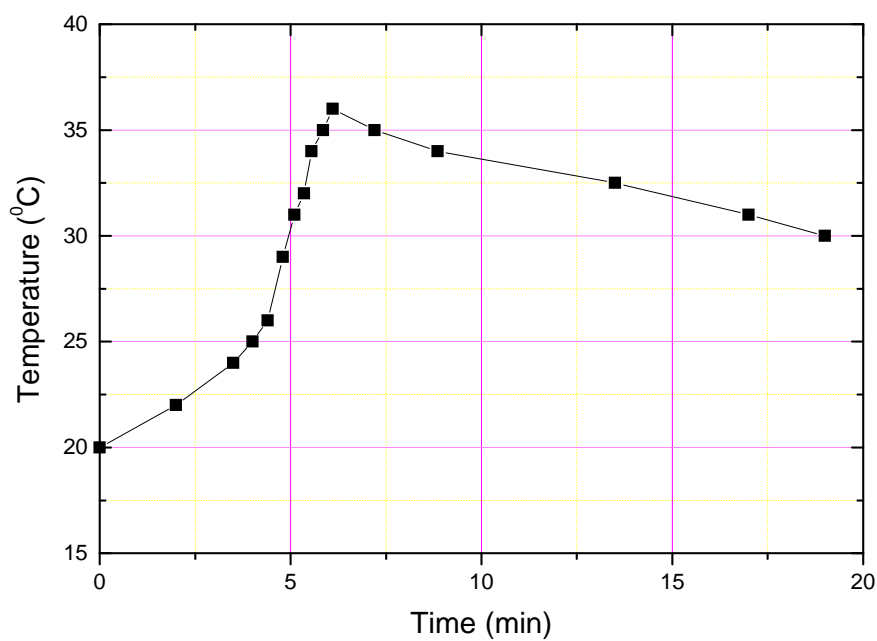

Fig.3: Temperature profile of the polymerization of aniline hydrochloride oxidized with ammonium peroxydisulfate at room temperature.

- **Preparation of polyaniline base (PANI base)**

The PANI salt prepared using the above procedure was dedoped using large excess of 0.5 M $NH_4OH$ to obtain the corresponding base. The $NH_4OH$ solution containing polyaniline-HCl salt was kept overnight under vigorous stirring and then filtered. The residue was washed with 500 mL of 0.5M $NH_4OH$. The deep blue polyaniline emeraldine base was dried for 72 hrs under vacuum.

- **Preparation of polyaniline salt doped with p-toluene sulfonic acid (PANI-pTSA)**

Obtained PANI base was protonated with p-toluene sulfonic acid 1.5 M solution in water. The protonation was achieved by extended mixing of PANI base and the acid solution for 24 hrs. The solution was filtered and a dark green emeraldine salt was obtained.

## 2.2. Structural characterization

FTIR spectra of the polyaniline samples have been taken using Bruker vector22 FTIR spectrophotometer. Pellets of PANI in KBr was prepared to register the IR spectra.

UV-visible spectra of polyaniline hydrochloride solutions in chloroform and m-cresol have been taken using Jasco-V530 UV-visible spectrophotometer.

A computer interface X-ray powder diffractometer (Philips) with Cu K$\alpha$ radiation ($\lambda = 0.1542\,nm$) was used. The data collection was over the 2-theta range of $10^0$ to $70^0$ in steps of $0.02^0/\sec$.

## 2.3. Density measurements

Polymer pellets of 25 mm in diameter and 0.4-0.5 mm thick were prepared from powder using a cylindrical die at 60 MPa.

The densities of polymer pellets were measured using the Archimedes principle with a Sartorius electronic balance equipped with a density measurement kit.

## 2.4. Electrical conductivity of polyaniline

Conductivity of prepared polyaniline was measured at room temperatures by a collinear four-probe method on polymer pellets prepared as mentioned in 2.3. A high impedance current source (Keithley 220) is used to supply current through the outer two probes (1 and 4) ), Fig.4. The voltage across the inner two probes (2 and 3) is measured by an electrometer (Keithley 619). Probe spacing $s$ is about 1.6 mm. Sample conductivity is determined by the relation:

$$\sigma = \frac{I}{2\pi sFV}$$

Where $F$ is a correction factor that depends on sample geometry and takes account of edge and thickness effects.

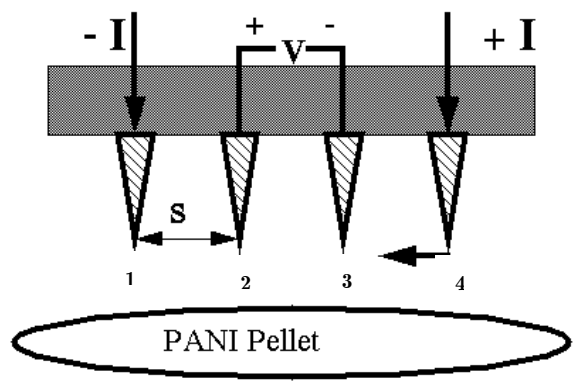

Fig.4: Schematic of 4-point probe configuration.

The influence of doping on the electrical conductivity of PANI was investigated by measuring the complex impedance of the polymer using an HP 4192A impedance analyzer in the range 10 kHz-13 MHz.

3. **Results and discussions**

3.1. **Preparation conditions**

Polyaniline of higher molar mass is produced at lower reaction temperatures, but the electrical properties of the polymer are not greatly influenced by molar mass, Fig.5. So, as we are interested essentially in electrical conductivity of PANI, we have selected ambient temperature for the present work. On the other hand washing the PANI precipitate with 0.2 M HCl removes residual monomer, oxidant, and its decomposition products. The treatment with hydrochloric acid solution provides a more uniform protonation of PANI with chloride counterions, although some of the sulfate or hydrogenosulfate anions from the decomposition of peroxydisulfate also participate as counterions. A final washing with acetone removes low-molecular oligomers. It also prevents the formation of agglomerates of PANI during drying, and the powder is obtained as a fine powder [4].

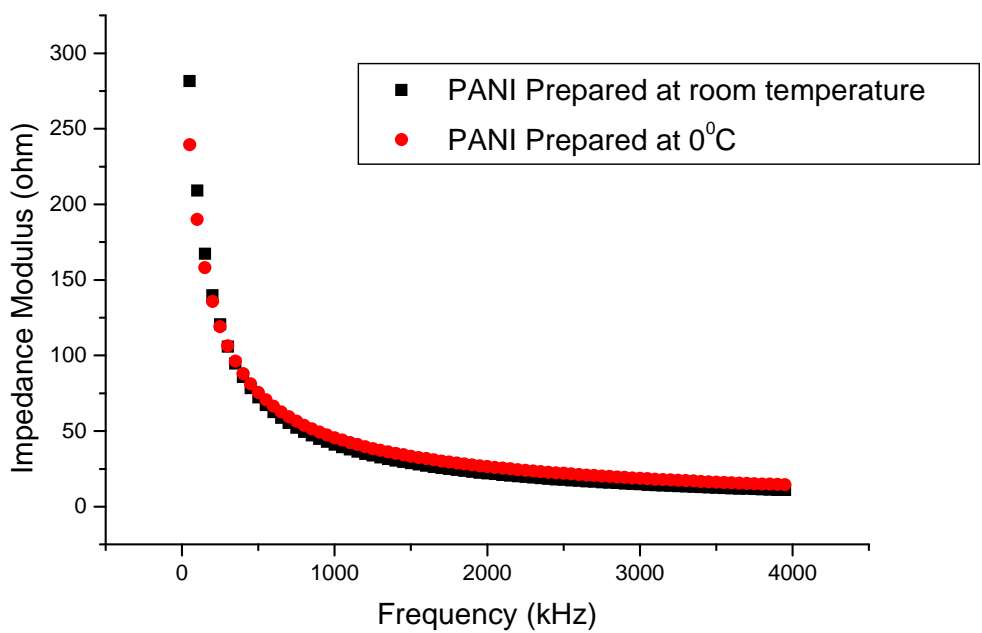

Fig.5: Effect of preparation temperature on the electrical properties of PANI.

### 3.2. Structural characterization

Fig.6 shows the IR spectra of PANI salt and PANI base. The peaks for the polyaniline base and salt appear at the same region and with similar intensities except for marginal differences. IR spectrum of the polyaniline base shows six principal absorptions at 1590, 1508, 1304, 1210, 1145 and 831 cm$^{-1}$. The peaks at 1590 and 1508 cm$^{-1}$ are assigned to C-C ring stretching vibrations. The peaks at 1304 and 1210 cm$^{-1}$ correspond to N-H bending and the symmetric component of the C-C (or C-N) stretching modes. The bands at 1145 and 831 cm$^{-1}$ can be attributed to the in-plane and out-of-plane C-H bending modes, respectively. The corresponding peaks for the polyaniline salt appear at 1560, 1482, 1306, 1245.9, 1148 and 814 cm$^{-1}$ respectively [5].

The strong band at 1145 cm-1 in the base is much intense and broader in the salt spectrum. In addition to the above peaks, the spectrum of the polyaniline salt exhibits new peaks around 3220, 1653 and 684 cm$^{-1}$. The peak at 3320 cm$^{-1}$ could be attributed as due to $NH_2$ stretching mode and the 1653 cm$^{-1}$ band to the $NH_2$ bending vibration while the 684 cm$^{-1}$ is assigned as due to $NH_2$ wagging.

The band characteristic of conducting protonated form is observed at 1246 cm$^{-1}$ [3,6]. It has been interpreted as originating from bi-polaron structure, related to C-N stretching vibration. The peak at $1580\,\text{cm}^{-1}$ confirms the presence of a protonated imine function. The C-Cl stretching peak arises in the region $590-700\,\text{cm}^{-1}$ [5].

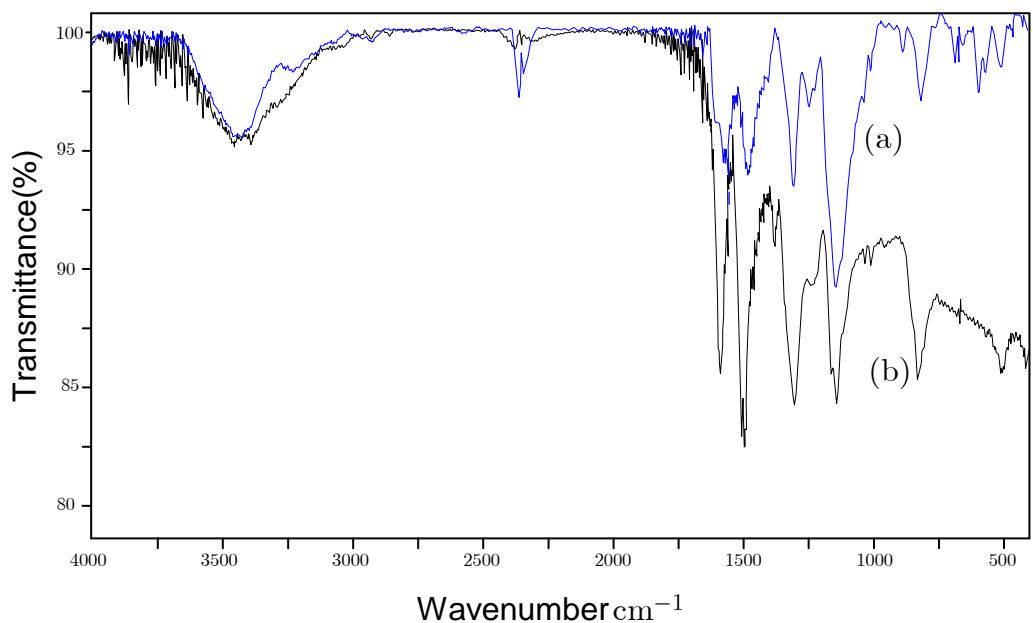

Fig.6: FTIR spectra of: (a) PANI hydrochloride and (b) PANI base.

Fig.7 and Fig.8 show the UV-visible spectra of PANI solution in chloroform and m-cresol respectively.

The UV-visible spectrum of PANI solution in chloroform has three absorption peaks ($<300\,\text{nm}, \sim 400\,\text{nm}, 550\,\text{nm}$), while the UV-visible spectrum of PANI solution in m-cresol has an absorption peak at $\sim 440\,\text{nm}$ and a steadily increasing "free carrier tail" from $1000\,\text{nm}$ to the IR region. We have to notice that the free carrier tail in the IR region is characteristics of metallic conductive materials. The polymer solution in m-cresol ($\sim 2\%\,\text{w/w}$) was deep green in color and viscous. While the polymer solution in chloroform ($\sim 2\%\,\text{w/w}$) was green in color and was not viscous.

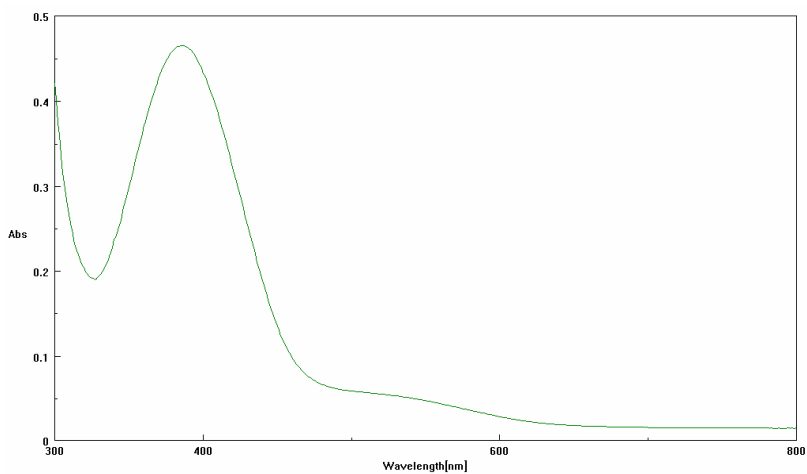

Fig.7: UV-visible spectrum of PANI in chloroform.

Based on the UV-visible spectra, we now know that polymer chains have different electronic structures in chloroform and m-cresol. This difference in electronic structures can be attributed only to their difference in conformational structures.

Specifically, polymer chains of PANI have a more extended conformation and hence a longer conjugation length in m-cresol than in chloroform.

The dispersion of the polaron band is, in particular very sensitive to the geometric arrangement of polymer backbone because the polaron band is formed via overlap between polarons of adjacent tetrameric units. In chloroform, the polymer chains have a random coil conformation. The polarons of individual tetrameric units are isolated from each other because of the twist defects between aromatic rings [1]. Therefore, the polaron band has little dispersion in energy, Fig 9 (a). The three absorption peaks in the UV-visible spectrum can be assigned as the transitions from $\pi$ band to $\pi^*$ band, polaron band to $\pi^*$ band, and $\pi$ band to polaron band respectively.

In m-cresol, the polymer chains of PANI have an extended conformation, in which the twist defects between aromatic rings are removed. The interaction between the adjacent polarons, therefore, becomes stronger, and the polaron band becomes more dispersed in energy (that is, more delocalized), Fig. 9 (b). As a result, the absorption peak at 550 nm (that is, the isolated polaron band) associated with the random coil conformation is replaced by intraband transitions within the half-filled polaron band (shown as the free carrier tail in the near-IR region). At the same time, the transition between $\pi$ band to $\pi^*$ band (<300 nm) becomes very weak and finally disappears, since the band gap between $\pi$ band and the polaron band has been eliminated.

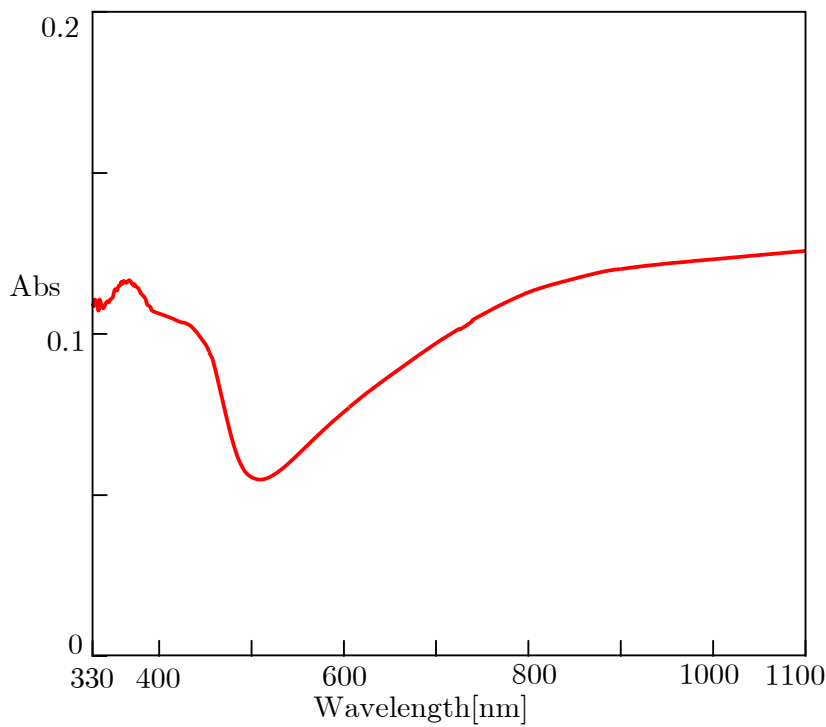

Fig.8: UV-visible spectrum of PANI in m-cresol.

In fact, the bulk properties of a polymeric material -for example, viscosity, crystallinity, elasticity, and the glass/melting point- are determined by the conformation of the polymer chains. For a conjugated polymer, the conjugation length, the electronic structure, and the electrical/optical properties of this polymer are strongly related to the conformational structure of the polymer chain.

Polyaniline emeraldine salt PANI is a polyelectrolyte in which the polymer backbone is positively charged with negative counterions sitting in the vicinity of polymer chains. The conformation of PANI in a solution depends on the nature of the solvent, the doping level (that is, the charge density on the polymer backbone), and the ionic strength of the solute ion.

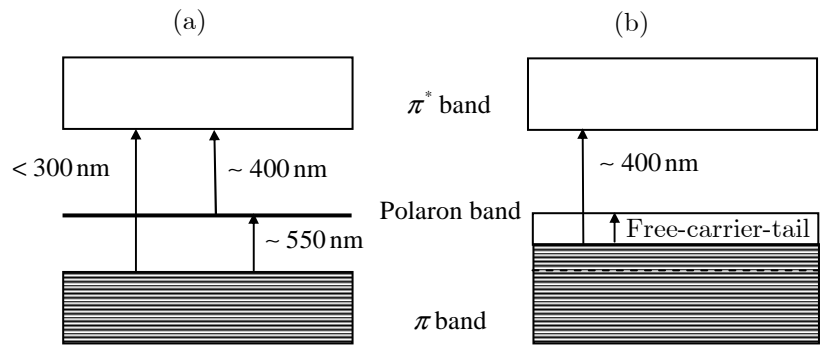

Fig.9: Electronic transitions of PANI solution in chloroform (a), and in m-cresol (2).

The X-ray diffraction pattern, Fig.10 indicates that the chains are strongly disordered, The PANI shows only a broad amorphous scattering around $2\theta = 25°$ [1,7].

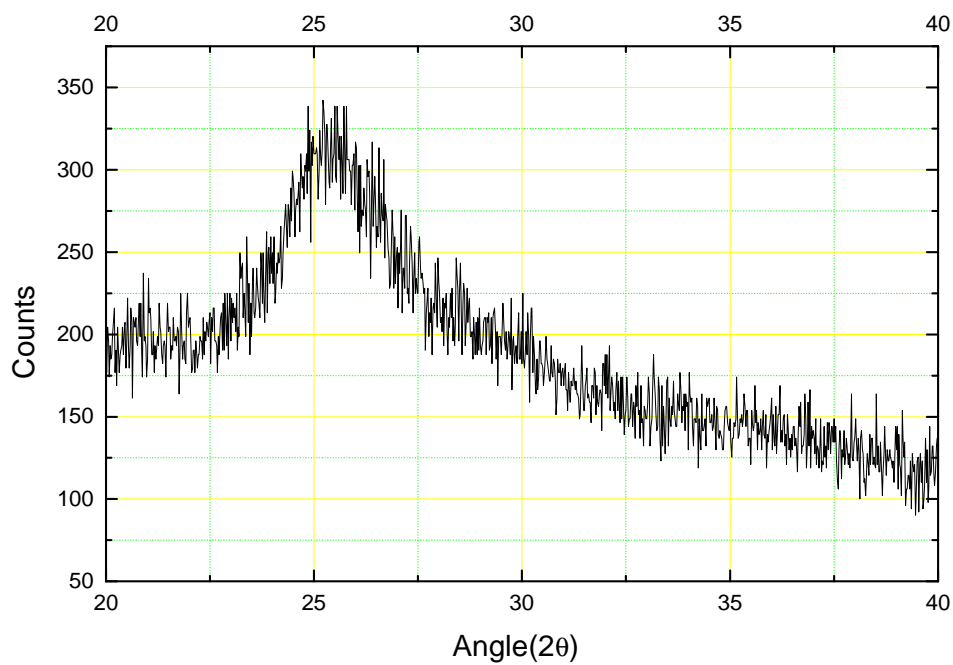

Fig.10: PANI hydrochloride X-ray diffraction pattern.

The Archimedes method was used to determine the polymer density by weighing the pellets in air and in n-heptan at 20°C, Tab.1. We remark that the density decreases for PANI base compared to the density of PANI hydrochloride, as a result of the substitution of chloride ions by hydroxyl group. While the density rises for polyaniline doped with p-toluene sulfonic acid compared to the density of PANI hydrochloride. This is due to the substitution of chloride ions by sulfonate groups. This result is close to that reported for polyaniline sulfate 1.43 [8].

| Density at 20°C | |
|---|---|
| Polyaniline hydrochloride | 1.348 ± 0.010 |
| Polyaniline base | 1.242 ± 0.019 |
| Polyaniline p-toluene sulfonic acid | 1.440 ± 0.001 |

Tab.1:Density of Polyaniline hydrochloride, Polyaniline base and Polyaniline p-toluene sulfonic acid at 20°C.

## 3.3. Electrical Conductivity

The electrical conductivity of PANI pellets measured using a four-probe method was 1.7 S/cm. So the green protonated emeraldine has conductivity on a semiconductor level [4]. This value is many orders of magnitude higher than that of common polymer $10^{-9}$ S/cm, but lower than that of typical metals $> 10^{4}$ S/cm.

It should be noted that the electrical conductivity of PANI augments nearly three times as we doped PANI with pTSA.

Figure 11 represents the impedance modulus of PANI hydrochloride and PANI-base pellets. Whereas Figure 12 represents the impedance modulus of PANI hydrochloride and PANI-pTSA pellets. We clearly constat that the PANI-base is insulator. While the PANI-pTSA pellet is better conductor than the PANI hydrochloride one.

In fact, a bipolaron is formed by two-electron oxidation of small segment of the polymer. However, this bipolaron can easily lose protons from NH groups where the positive charge is localized, as shown below:

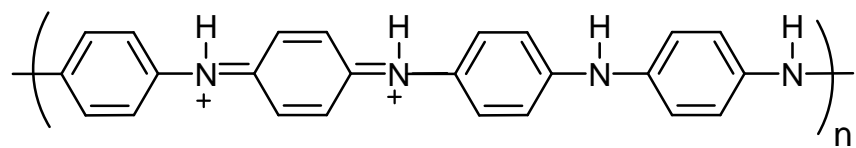

PANI-hydrochloride, Emeraldine Salt , green, partially oxidized, protonated conducted

The resulting quinoid form (called "emeraldine base" form is insulating:

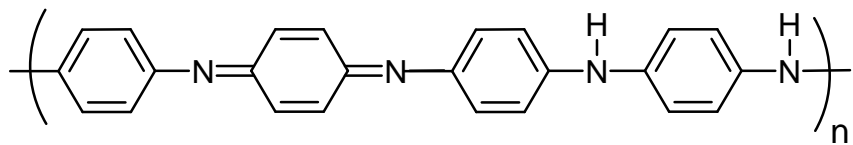

PANI-base- Emeraldine Base, partially oxidized, insulating.

When the emeraldine base is protonated by acid (pTSA by example), each nitrogen again has a proton attached, and the bipolaron can move along the chain without the need for protons to hop as well. This form is a good electronic conductor. The polymer can be switched back and forth between conducting and insulating forms by simply changing the pH.

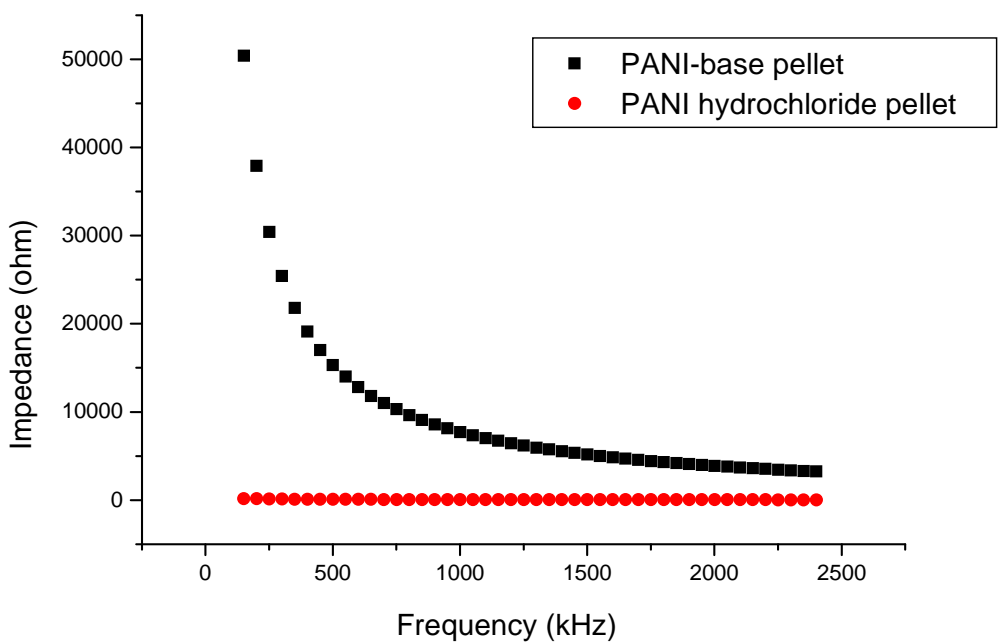

Fig.11: Impedance modulus of PANI hydrochloride and PANI-base pellets.

We notice that the conductivity of a PANI pellet prepared from a solution in m-cresol is better than that of a PANI pellet prepared from a solution in chloroform Figure 13. This difference in electrical properties can be attributed only to their difference in conformational structures. In the case of m-cresol, we have a polaron band formed via overlap between polarons of adjacent tetrameric units. While in chloroform The polarons of individual tetrameric units are isolated from each other as it is discussed in 3.2.

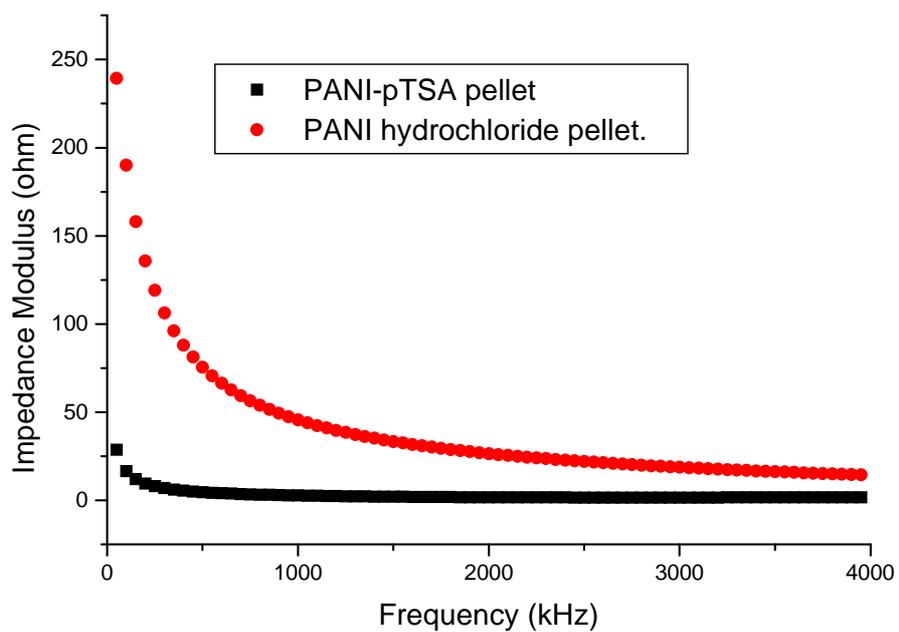

Fig.12: Impedance modulus of PANI hydrochloride and PANI-pTSA pellets.

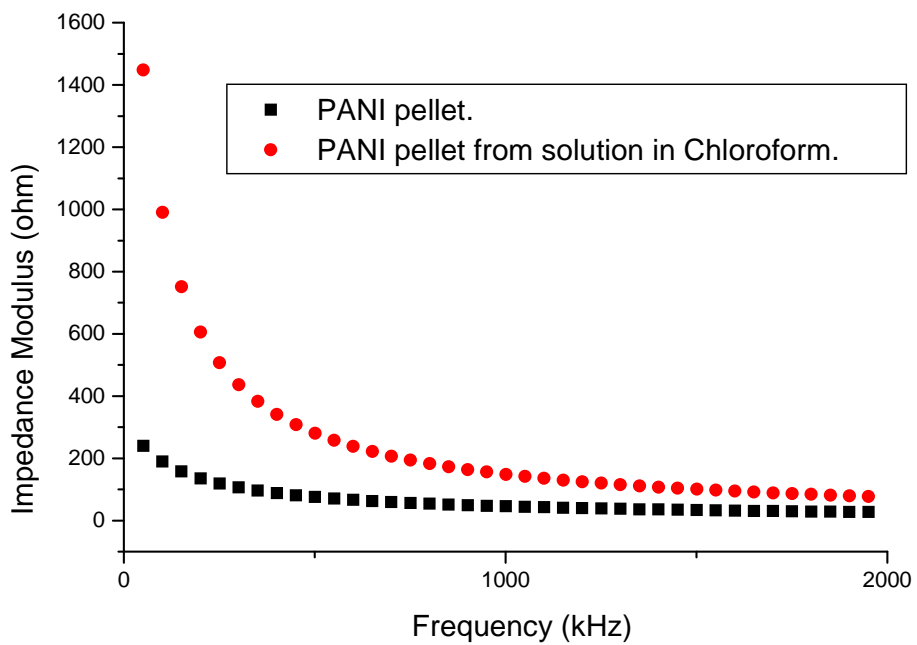

Fig.13: Impedance modulus of PANI hydrochloride pellets from m-cresol and from chlorofom.

4. **Summary**

The polyaniline has been successfully synthesized by chemical oxidation using anilinium hydrochloride as a monomer. The handling of solid aniline salt is preferred to liquid aniline from the point of view of toxic hazards. We illustrated clearly why the conducting form of polyaniline is called "emeraldine" and the reversible switching between conductive and nonconductive forms via protonation/deprotonation. According to electrical measurements the conductivity of the polymer is in good agreement with the expected values as a semiconductor material.

**References:**

[1]. D. L. Wise, G. E. Wnek, D. J. Trantolo, T. M. Cooper, and J. D. Gresser, Marcel Dekker, Inc., New York, 1996.

[2]. A. Malinauskas, Polymer, 42 (2001)3957.

[3]. S. Chakane, P. Likhite, S. Jain, and S.V. Bhoraskar, Transactions of the SAEST, No. 1, 37(2002)35.

[4]. J. Stejskal, R.G. Gilbert, Pure Appl. Chem., No. 5, 74 (2002)857.

[5]. P. S. Rao, J. Anand, S. Palaniappan and D. N. Sathyanarayana, European Polymer Journal, 36 (2000) 915-921.

[6]. A.M.P. Hussain and A. Kumar, Bull. Mater. Sci., 26 (2003)329.

[7]. M. G. Han and S.S. Im, Journal of applied Polymer Science, 71(1999)2169.

[8]. J. Stejskal, I. Sapurina, M. Trchova, J. Prokes, I. Krivka, E. Tobolkova, Macromolecules, 31 (1998) 2218.